\documentclass[12pt]{iopart}

\usepackage{graphicx}
\usepackage{epsfig}

\newcommand{\eqref}[1]{(\ref{#1})}

\newcommand \M {\ensuremath{{\cal M\,}}}
\newcommand \Msol {\ensuremath{{\rm M}_\odot}}
\newcommand \Hz {\ensuremath{\mathrm{Hz}}}

\begin{document}

\title[Impact of LISA's low frequency sensitivity]{Impact of LISA's low frequency sensitivity on observations of massive black hole mergers}

\date{\today}

\author{J Baker and J Centrella}

\address{NASA-Goddard Space Flight Center, Greenbelt, MD 20771}

\begin{abstract}
LISA will be able to detect gravitational waves from inspiralling
massive black hole (MBH) binaries out to redshifts $z > 10$. If the binary
masses and luminosity distances can be extracted from the LISA
data stream, this information can be used to reveal the merger
history of MBH binaries and their host galaxies in the evolving universe.
Since this parameter extraction generally requires that LISA observe
the inspiral for a significant fraction of its yearly orbit, carrying
out this program requires adequate sensitivity at low frequencies,
$f < 10^{-4}$ Hz.  Using several candidate low frequency sensitivities,
we examine LISA's potential for characterizing MBH binary coalescences
at redshifts $z > 1$.

\end{abstract}

\ead{Joan.Centrella@NASA.gov}

\submitto{\CQG}

\maketitle

\section{Introduction}
\label{sec:intro}

The final coalescence of an MBH binary is a
strong source of low frequency gravitational waves for LISA. Current
studies in $\Lambda$CDM hierarchical models predict $\sim 100$ such
events per year out to redshifts $z > 10$ 
\cite{Sesana:2004gf,SesanaProc,Wyithe:2003},
although the rates could be $\sim 1$ event per year or lower
\cite{Haehnelt:2003wa,HaehneltCarnegie}.
Observations of these waves can potentially be used to map the
distribution of MBH binaries in the evolving universe, yielding key
astrophysical information on the merger history of the MBHs,
their host galaxies and the development of structure in the
universe.  Success in carrying out this program requires that source
parameters such as the binary masses, spins, and luminosity distance
be extracted with sufficient accuracy from LISA's data. Obtaining
good accuracy in extracting these parameters 
is generally expected to depend on observing
the sources throughout a significant portion of LISA's yearly orbit
about the sun. For many MBH sources, this requires adequate
sensitivity at low frequencies, $f < 10^{-4} {\rm Hz}$.

LISA will observe the final stage of MBH evolution, which is driven
by gravitational radiation reaction and proceeds in three phases: an
adiabatic inspiral, followed by a dynamical merger and a final
ringdown. The relatively slow inspiral produces chirp waveforms,
sinusoids increasing in frequency and amplitude as the orbital
period shrinks; throughout most of this phase, the waveforms can be
computed using post-Newtonian techniques \cite{Blanchet:2002av}. 
By the time the BHs
reach a separation $\sim 6M$ (we set $c = G = 1$ for simplicity),
the evolution takes on a dynamical character as the BHs plunge
towards each other and merge into a single, highly distorted BH.
Calculating the waveforms from this stage requires full 3-D
numerical relativity; while recent progress towards this goal is
encouraging ({\em e.g.}, \cite{Bruegmann:2003aw}), 
reliable merger waveforms are not yet available.
Finally, during the ringdown, the distorted and vibrating remnant
radiates away its nonaxisymmetric modes and evolves into a quiescent
Kerr BH. These ringdown waves are damped sinusoids and can be
computed using perturbative methods \cite{Leaver:1985ax,Echeverria:1989hg}.

In this paper, we focus on LISA's potential for successfully
characterizing MBH binary coalescences at redshifts $z > 1$ by
observing inspiral waves and using parameter extraction.  For
several candidate low frequency sensitivities, we map out the
science reach of LISA for the frequency range $10^{-6} {\rm Hz} < f
< 10^{-4} {\rm Hz}$.

\section{Observing MBH Inspiral with LISA}
\label{sec:ObsMBH}
 Signals from inspiralling MBH binaries
  can be found in the LISA data stream
using matched filtering based on templates calculated from
post-Newtonian waveforms. The expected signal-to-noise ratio (SNR)
for observations of chirping sources based on matched filtering is
 \cite{Flanagan:1998sx}
\begin{equation}\label{eq:SNR}
({\rm SNR})^2= \int_0^{\infty}\frac{df}{f}\frac{h_c^2(f)}{\langle f
S_h(f)\rangle},
\end{equation}
where $S_h(f)$ is the instrumental strain noise spectrum (in
$\Hz^{-1}$) and the angle brackets denote sky averaging. The
characteristic strain of the signal $h_c$ can be written, after
averaging over orientations, 
\begin{equation}
h_c = \frac{\sqrt{2}(1+z)}{\pi D_L(z)} \sqrt{\frac{dE_e}{df_e}},
\label{hc1}
\end{equation}
where $dE_e/df_e$ is the total energy per unit frequency carried
away by the gravitational waves and the subscript "e" refers to an
``emitted'' quantity
calculated in the source's reference frame.  LISA measures
a redshifted frequency $f = f_e/(1+z)$. We take the luminosity
distance--redshift relation, $D_L(z)$, as determined by observed
cosmological parameters, $\Omega_\Lambda=1-\Omega_M=0.73$ and
$h=0.71$ based on recent WMAP data \cite{Spergel:2003cb}. This
construction allows a sky and orientation averaged estimate of
LISA's probable level of sensitivity to a system's gravitational
wave signal to be produced as a function of frequency simply in
terms of the overall power spectrum of the gravitational radiation
energy released.

We assume circularized orbits and quadrupolar inspiral waveforms,
leaving out the effects of spin and keeping only the lowest order
(Newtonian) terms.  We consider MBH binary systems to have mass
ratios ranging from $m_2/m_1 > 0.3$ for major mergers down 
to $m_2/m_1 > 10^{-2}$ for minor mergers; 
for smaller mass ratios, the systems approach
a compact object capture scenario in which the waveforms typically
have a different character \cite{Hughes:2000ss}.
The radiation power
spectrum for an inspiralling MBH binary can be obtained from the
binary system energy $E_e = (1/2)\mu (\pi M f_e)^{2/3}$, where $M =
m_1 + m_2$ is the total mass and $\mu = m_1 m_2/M$ is the reduced mass
of the binary. Then equation \eref{hc1} gives
the characteristic strain
\begin{equation}
h_{\rm c} = 6.5 \times 10^{-17} \left(\frac{D(z)}{1 \;{\rm  Gpc}}\right)^{-1}
 \left( \frac{(1+z)\M}{10^6 \Msol} \right)^{5/6} \left(\frac{f}{10^{-4} \rm Hz}\right)^{-1/6}
\label{hc}
\end{equation}
for a binary with chirp mass $\M = (M^2 \mu^3)^{1/5}$ at redshift
$z$.

Because the quality of the observations depends on the length of
time that the system's signal remains in LISA's band, we also
require an expression for the binary frequency as a function of
time. To lowest order this gives \cite{Blanchet:2001ax} $(\pi M f_e)^{2/3} =
\frac{1}{4} \tau^{-1/4}$, where $\tau = \mu (t_c - t)/(5M^2)$, $t_c$
is the time of coalescence (when $f_e \rightarrow \infty$), and $t_c
- t$ is the amount of time the binary can be observed locally before
coalescence.  Thus
\begin{equation}
f = 5.4 \times 10^{-5}\, {\rm Hz} \left ( \frac{\M (1+z)}{10^6 \Msol} \right )^{-5/8} \left (
\frac{t_{\rm obs}}{6\;{\rm mos.}} \right )^{-3/8}, \label{foft}
\end{equation}
where $t_{\rm obs} = (1+z)(t_c - t)$ is the amount of time that the
binary can be observed by LISA.

For the purpose of this analysis it turns out to be adequate to keep
only lowest order terms. Specifying that $\tau>40$ ensures that these 
lowest order expressions for $f(\tau)$ and $h_c(f(\tau))$ (see equation
\eref{eqn:strain})
 agree with the higher-order post-Newtonian expressions 
to within $\sim 25\%$. 
For major mergers ($m_2/m_1 >0.3$) this condition on $\tau$ is met when
$\M(1+z)<4\times10^{10} \Msol$. For smaller mass ratios the applicable range 
is reduced, with the requirement $\M(1+z)<4\times10^8 \Msol$ at 
$m_2/m_1 >0.01$.

\begin{figure}[t]
\includegraphics[scale=.55,angle=-90]{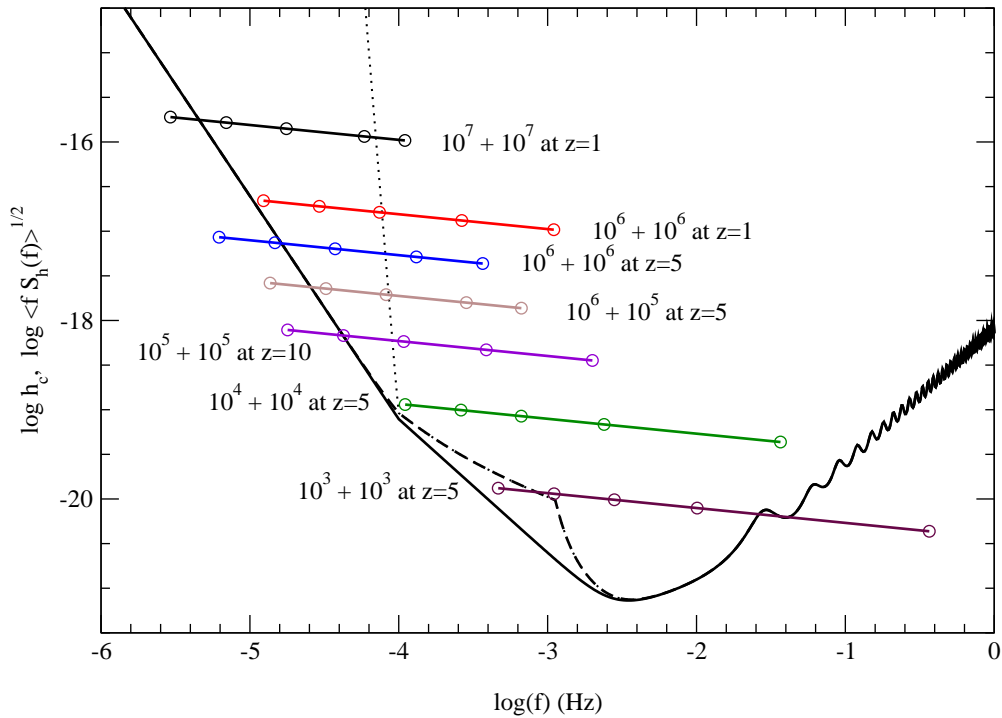}
\caption[]{Characteristic strain $h_c$ 
evolution curves for several MBH binaries (masses given
in \Msol) at various redshifts.} \label{fig:traces}
\end{figure}
Figure \ref{fig:traces} shows the evolution of inspiralling MBH binaries at various
redshifts relative to LISA's sensitivity. 
The thick solid line shows the baseline design sensitivity as can be
obtained from the LISA Sensitivity Curve Generator \cite{LISASenGen}
 with an extension as $\sqrt{S_h(f)}\propto f^{-2}$ to frequencies 
below $10^{-4}$Hz based on the optimistic assumption of white acceleration
noise. The dotted line represents a
pessimistic scenario with a nearly vertical 
($\sqrt{S_h(f)}\propto f^{-20}$) cut-off in sensitivity
below $10^{-4}$Hz.  The thick dashed line shows the expected
confusion noise due to galactic binaries using the expressions in Barack and 
Cutler \cite{Barack:2003fp}, which are compiled from 
\cite{Farmer:2003pa,Nelemans:2001}. Starting at the low frequency end,
the symbols on each
binary evolution track denote the system at 10 years, 1 year, 1
month, 1 day before the dynamical merger, and at the onset of merger.
The merger itself, as with the subsequent ringdown, produces radiation
at higher frequencies.

The signal can be said to come into band when $h_c(f)\ge\langle f
S_h(f)\rangle^{1/2}$. Even for the very pessimistic low frequency
scenario (dotted line in figure~\ref{fig:traces}), 
all of these systems should be detectable by LISA since
they have characteristic strains $h_c$ significantly above the
sensitivity curve. As can be 
seen in equation \eref{eq:SNR},
the height of $h_c$ above the sensitivity
curve provides an indication of the signal-to-noise ratio. 
However, to extract  the binary parameters, and in particular the
luminosity distance, the source must be observed by LISA over an
extended period of time.  For many sources, particularly those at
higher $z$, this requires good sensitivity below $10^{-4}$Hz.

\section{Parameter Estimation}
\label{sec:param}

LISA is not a pointed instrument, but rather an all-sky monitor with
a predominantly 
quadrupolar antenna pattern.  The orbital motion of LISA around
the sun and the yearly rotation of the spacecraft constellation
around the normal to the detector plane (which is tilted by
$60^{\circ}$ with respect to the ecliptic) induce modulations of
the incident gravitational waveforms that encode the source's sky
position and orientation. LISA measures both polarization components
of all incident gravitational waves simultaneously, so its
astronomical data will consist of essentially 
two time series.  All the physical
properties of a source, including its position and distance, must be
extracted from these data streams.

This extraction process is complicated by the fact that the source
parameters are entangled.  For an MBH binary, the redshifted
component masses, $(1+z)m_1$ and $(1+z)m_2$, can be measured by
tracking the phase of the inspiral waveforms and matching the
results to templates.  The overall amplitudes of the gravitational
waves depend on the luminosity distance $D(z)$, the chirp mass $\M$,
and the orientation and sky position of the binary. Significant
motion-induced modulations of the gravitational waveforms are needed
to determine the latter two parameters well.  Good observations,
particularly those allowing the redshift $z$ to be extracted along
with the masses, thus require the binaries to be within the band of
sensitivity for a significant fraction of LISA's yearly orbit.

 Cutler \cite{Cutler:1998ta} produced the first detailed analysis of how
precisely source parameters may be estimated from LISA observations.
Hughes \cite{Hughes:2001ya} expanded on this work, providing detailed
estimates of LISA's precision in determining MBH binary positions
and distances with the assumption of poor low frequency performance.
Hughes and Holz \cite{Hughes:2003ty} considered the problem of how to
apply LISA observations of MBH binary systems in cosmological
studies. Specifically comparing the effect of a cutoff in LISA
sensitivity at $10^{-4}\Hz$ with that of a lower cutoff at $3 \times
10^{-5}\Hz$, they found that the lowered cutoff provides a factor of
ten improvement in distance measurement precision for an MBH binary
with $m_1 = m_2 = 10^6M_\odot$. Based on their studies Hughes and
Holz proposed a rule-of-thumb, that a good
measurement of source orientation and, by association, distance
requires that LISA moves through one radian of its orbit,
corresponding to approximately two months of observation in band.

These studies did not take into account the imprint of spin-induced
precession of the orbital plane on the waveforms.
Vecchio \cite{Vecchio:2003tn} recently showed that consideration of
these precession effects in the data analysis can dramatically
improve LISA's parameter estimates. Rather than imposing a low
frequency cutoff, he assumed that LISA observes the entire final
year of the MBH binary inspiral. For two inspiralling, rapidly
rotating $10^6 M_{\odot}$ MBHs at $z = 1$, Vecchio finds that the
luminosity  distances can typically be estimated to within $\sim 1\%$
and the masses to within $\sim 0.01\%$, while the location of the source on
the sky is typically determined within an area of about a square degree.
This requires the signal
to be visible to LISA for $\sim 6$ months; longer observation times
provide only marginal improvements in the accuracy of these
parameters \cite{Vecchio2004Proc}. Below we will
apply this 6 month criterion as a tool to estimate the science reach
implied by various LISA sensitivity curves.

\section{Low frequency sensitivity for LISA}
\label{sec:candidates}

In figure~\ref{fig:traces} we provided two sensitivity curves,
representing potential realizations of LISA's low frequency strain
noise power spectrum $S_h(f)$.  Here we consider several other plausible 
low frequency sensitivity realizations.

The ``baseline'' sensitivity curve in figure~\ref{fig:traces} should be
considered extremely optimistic because it makes no allowance for likely 
additional sources of acceleration noise which enter at low frequencies.  
Specifically including some of these
effects, Bender \cite{Bender:2003uv} has proposed a more realistic
sensitivity curve which might be expected without very costly enhancements
of the baseline design.  His curve extends as $\sqrt{S_h(f)} \propto
f^{-2.5}$ from $10^{-4}\Hz$ down to $10^{-5}\Hz$, and then as
$\sqrt{S_h(f)} \propto f^{-3}$ out to $3 \times 10^{-6}\Hz$. We call
this case ``Bender'', adding, in this and all cases, an extension
$\sqrt{S_h(f)} \propto f^{-6}$ below $3 \times 10^{-6}\Hz$.  This
curve just meets a proposed LISA science requirement at $10^{-5}\Hz$
\cite{TomsLSR}. For comparison, we consider a curve, ``Relaxed'',
which might correspond to a more relaxed requirement extending out
directly as $\sqrt{S_h(f)} \propto f^{-3.5}$. The strongly pessimistic
extension with a noise ``Wall'' below $10^{-4}$ is similar to the low frequency sensitivity considered in \cite{Hughes:2001ya}.
Because the baseline design is not guaranteed to be realized even above 
$10^{-4}\Hz$, we also consider a case with relaxed sensitivity across the band.  This case corresponds to a conceivably ``de-scoped'' version of LISA, degraded by a factor of ten from the Relaxed case.

\section{Science Reach for MBH Binary Inspirals}
\label{sec:reach}

Uncovering the merger history of MBHs in the evolving universe using
gravitational waves requires that LISA observations provide the
masses, sky positions, and redshifts of these sources.  Quantifying
this discovery space requires that we specify a threshold criterion
for a sufficently informative observation.  Based on the discussion in
section~\ref{sec:param}, we apply the rule-of-thumb that LISA must
observe the source for $t_{\rm obs} = 6$ months in band. Otherwise,
recent studies of parameter estimation accuracy suggest that we should
not expect a good measurement of the luminosity distance and position
on the sky.  It should be noted that incomplete information, such as
the redshifted chirp mass $(1+z)\M$, may nonetheless be determinable
for systems which fail our criterion.

We now proceed to map this discovery space parametrized in terms of
the chirp mass \M and the redshift $z$. At a fixed time $t_{\rm obs}$
prior to coalescence, the system's signal is characterized by
frequency $f(t_{\rm obs},\M,z)$ given by equation \eref{foft}.  We also
need the characteristic strain at this frequency, at time
$t_{\rm obs}$ before coalescence.  To obtain this, we eliminate \M in
equation \eref{hc}; this gives

\begin{eqnarray}\label{eqn:strain}
h_c(t_{\rm obs},z,f) = 2.8\times10^{-17}
\left(\frac{D_L(z)}{1 \;{\rm Gpc}}\right)^{-1}\left(\frac{t_{\rm obs}}{6\;
{\rm mos.}}\right)^{-1/2}\left(\frac{f}{10^{-4}\Hz}\right)^{-3/2}.
\end{eqnarray}
If, at $t_{\rm obs}=6$ months, this strain is larger than LISA's
annually averaged unitless strain noise at this frequency, 
{\em i.e.}  above the
sensitivity curve $h_c > \langle f
S_h(f)\rangle^{1/2}$, then a system of the specified chirp mass and
redshift will be identified as being ``in band'' for at least 6 months,
thus meeting the criterion for a minimum adequate observation.

\begin{figure}[t]
\includegraphics[scale=.55,angle=-90]{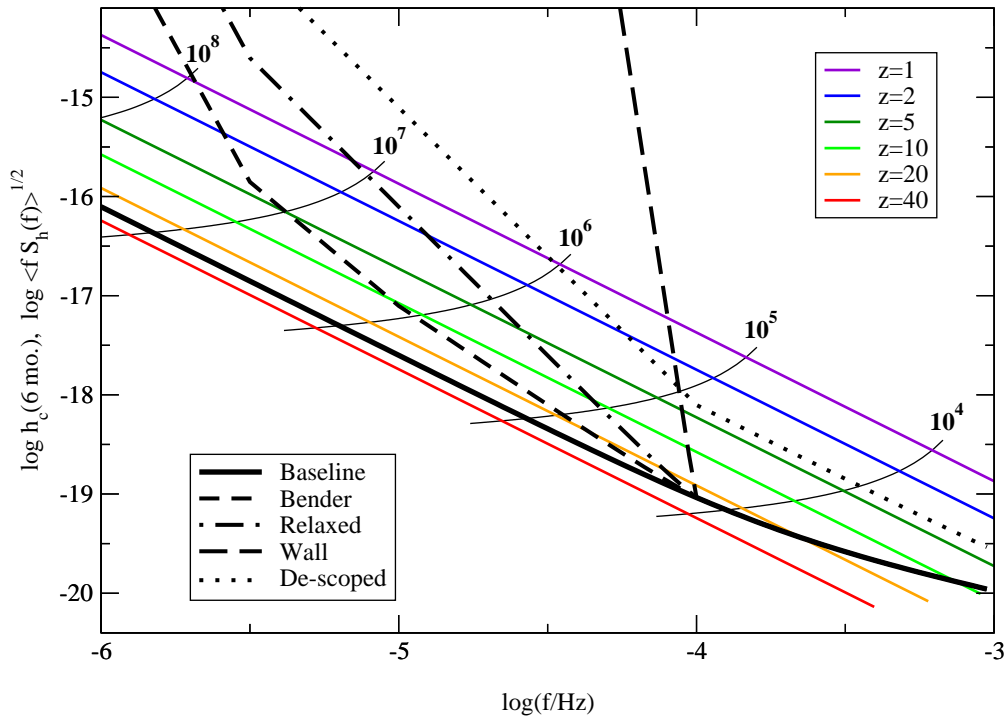}
\caption[]{Sources which can be adequately observed by LISA with
the various low frequency sensitivity curves discussed in 
Section~\ref{sec:candidates}.  Potential sources are
labeled by redshift $z$ and chirp mass $\M$ (in $\Msol$). 
The range of sources
falling above a particular sensitivity curve would be observable for
a minimum of six months by the corresponding realization of LISA.
The sensitivity curves shown represent a combination of instrumental 
noise and confusion noise from the astrophysical WD-WD background.
\label{fig:plots}}
\end{figure}

Plotting the values given by equations~\eref{hc} and  \eref{foft}  with
candidate realizations of LISA's unitless strain noise $\langle f
S_h(f)\rangle^{1/2}$ allows us to make a map, shown in
figure~\ref{fig:plots}, of which systems are accessible.
The sensitivity curves shown represent the instrumental sensitivity
models described in section~\ref{sec:candidates} added to a model
of binary white dwarf noise given in \cite{Barack:2003fp}. The
colored lines depict MBH binary systems at $t_{\rm obs} = 6$ months
prior to coalescence at specified redshifts ranging from $z=1$ to
$z=40$. These are crossed by
a set of thin solid dark curves denoting the chirp masses \M of systems
observed at frequency $f$ at time $t_{\rm obs} = 6$ months.  Together
these form a grid representing the variety of potentially observable
astrophysical systems. Regions of the grid which lie above a
particular sensitivity curve correspond to systems which can be
observed by that version of LISA for at least 6 months.

This figure indicates which systems we can expect the various
versions of LISA to observe effectively at large redshift. In
particular, it suggests that, for large redshift observations, LISA
will likely have a ``sweet spot'' near $10^{-4}\Hz$, and will be
most sensitive to systems of chirp mass around $\M \sim 10^4$-$10^5
\Msol$. If the baseline sensitivity is achieved at this frequency,
then the 6 month criterion suggests that LISA may be able to observe
such systems out to $z\sim20$.    Since the slope of the constant
redshift curves is precisely the same as the slope of the baseline
sensitivity curve dominated by white acceleration noise, this region
can be extended for as far as the baseline sensitivity curve can be
preserved, allowing larger systems to be observed to the same
redshift.

The ability to observe high-redshift systems with mass near $\M\sim
10^6\Msol$  will be determined by the
sensitivity near $10^{-5}\Hz$.  The difference between Bender's
proposal and the relaxed sensitivity curve is significant in this
range and implies a change by about a factor of three in the
redshift to which LISA will be able to see. Bender's
sensitivity allows observation to a redshift of $z \sim 10$.  A noise
wall above $3 \times 10^{5}\Hz$ might completely obscure the late
inspirals of these systems.  Likewise, observation of any systems
beyond $z=1$ with masses larger than $\M\sim10^7\Msol$ would require 
fair sensitivity below $10^{-5}\Hz$. De-scoping LISA as described in
Section~\ref{sec:candidates} would constrain observations to lower
redshifts, $z < 4$ for $\M \sim 10^4 \Msol - 10^5 \Msol$ and
$z < 1$ for $\M \sim 10^6 \Msol$.

\section{Discussion}
\label{sec:discuss}

We have considered the observability of MBH binary systems at
redshifts $z > 1$ by LISA for several possible low frequency
sensitivities, $f < 10^{-4}$Hz.  We distinguish between
{\em detection} of the binary, in which the LISA data stream would
contain a signal from these sources at good SNR ({\em cf.} 
figure~\ref{fig:traces}) and {\em observation}, in which the binary
parameters such as component masses, spins, luminosity distance
and sky position can be extracted accurately.  Assuming that successful
parameter extraction relies on using the motion-induced modulations
of the gravitational waveforms, we applied the rule that an MBH
binary system must be monitored by LISA for 6 months in band for
a good observation \cite{Vecchio2004Proc}.  
Figure~\ref{fig:plots} shows that good sensitivity,
given by the Bender curve, at $f \sim 10^{-5}$Hz will allow LISA to
observe MBH binaries with $\M \sim 10^6 \Msol$ out to $z \sim 10$.
Lower mass systems, $\M \sim 10^4 \Msol - 10^5 \Msol$, could be observed out
$z \sim 20$ if the sensitivity is near the baseline design at
$f \sim 10^{-4}$Hz.

We have considered only the lowest order quadrupolar waves from
MBH binary inspirals.  Including higher order multipole components
\cite{Hellings:2002si} could improve the parameter extraction,
lessening the amount of time needed for a good observation. Such a
relaxation of the 6 month-in-band rule would allow LISA to see MBH
binaries to even higher redshifts for the low frequency sensitivities
discussed here.  For example, using the Relaxed sensitivity curve,
equation~\eref{eqn:strain} shows that
LISA could observe $\M \sim 10^6 \Msol$ binaries out to 
$z \sim 8$ for a six-week-in band observation, should such
sources exist.  These considerations
deserve further study as the specification of LISA's design is completed.
Similarly, if monitoring only the final radiation burst is found
to allow an adequate estimation of the system parameters, or if 
incomplete observations that only allow determination of some 
parameters (such as $(1+z)\M$, for example) are considered adequate,
then many more systems could be deemed observable.  We are currently
investigating these issues.

Good low frequency sensitivity for LISA has considerable astrophysical
value.  Knowledge of the binary parameters will reveal the merger
history of MBHs and, by implication, their host structures as the
universe evolves.  Current models for MBH growth and cosmic structure
formation yield different predictions for the rate of MBH binary
mergers.  LISA observations of these systems will provide direct
dynamical information about the number of events at different
redshifts. Comparing these data with model predictions will discriminate
between different scenarios ({\em e.g.,} 
\cite{Sesana:2004gf,Wyithe:2003,Haehnelt:2003wa,HaehneltCarnegie}), 
shedding light on the relative
importance of mechanisms such as accretion and mergers for MBH
growth and structure formation.

\ack It is a pleasure to acknowledge stimulating discussions with
Scott Hughes, Alberto Vecchio, Tom Prince, and Richard Mushotzky.  
This work was 
supported in part by NASA through the LISA Project Mission Science Office.
Part of this work was performed while J.B. held a National
Research Coundil Associateship Award at the Goddard Space Flight
Center.

\section*{References}

\bibliographystyle{prsty}
\bibliography{white}

\end{document}